\documentclass[%
 reprint,
 amsmath,amssymb,
 aps,
]{revtex4-1}

\usepackage{graphicx}% Include figure files
\usepackage{dcolumn}% Align table columns on decimal point
\usepackage{bm}% bold math
\usepackage{amsfonts,color}
\begin{document}

%\preprint{APS/123-QED}

\title{Vibrational Spectrum of Granular Packings With Random Matrices}
%\\thanks\{A footnote to the article title\}%\

\author{Onuttom Narayan}
\affiliation{
Physics Department, University of California, Santa Cruz, CA 95064
}%\

\author{Harsh Mathur}
\affiliation{%
	Physics Department, Case Western Reserve University, Cleveland, OH 44106-7079
}%\

\date{\today}% It is always \\today, today,\
             %  but any date may be explicitly specified\

\begin{abstract}
The vibrational spectrum of granular packings can be used
as a signature of the jamming transition, with the density of states at zero 
frequency becoming non-zero at the transition. It has been proposed previously
that the vibrational spectrum of granular packings can be approximately obtained
from random matrix theory. Here we show that although the density of states predicted
by random matrix theory does not agree with certain aspects of dynamical
numerical simulations, the correlations of the density of states, 
which---in contrast to the density of states---are expected to be universal, do show
good agreement between dynamical numerical simulations of bead packs near the jamming point 
and the analytic predictions of the Laguerre orthogonal ensemble
of random matrices. At the same time, there is clear disagreement with the Gaussian orthogonal
ensemble. These findings establish that the Laguerre ensemble correctly reproduces the 
universal statistical properties of jammed granular matter and exclude the Gaussian 
orthogonal ensemble. We also present a random lattice model which is a physically motivated variant of
the random matrix ensemble. Numerical calculations reveal that this model
reproduces the known features of the vibrational density of states of granular
matter, while also retaining the correlation structure seen in the Laguerre
random matrix theory. We propose that the random lattice model can therefore
be applied the understand not only the spectrum but more general properties
of the vibration of bead packs including the spatial structure of modes both at the
jamming point and far from it. 

\end{abstract}

%\\keywords\{Suggested keywords\}%Use showkeys class option if keyword\
                              %display desired\
\maketitle

%\\tableofcontents\

\section{\label{sec:Intro} Introduction}
Granular materials are a class of systems which are out of equilibrium
and not easy to understand within the framework of standard statistical
mechanics.  For static assemblies, the distribution of forces~\cite{Coppersmith} and
the continuum limit~\cite{Cates} are difficult to obtain. This is because
interparticle contacts are very stiff: a slight compression of two
particles that are contact, by an amount that is much less than the
interparticle separation, gives rise to large forces. Added
complications are caused by the fact that, for noncohesive granular
matter, two particles in contact repel each other when they are
compressed, but do not attract each other when they are moved away
from each other and the contact is broken; that the repulsive force
between particles is not a linear function of their compression
when the compression is small;~\cite{MvHreview} and that there are frictional forces
between particles,~\cite{Bi} resulting in history dependent forces.  The
dynamic properties of granular matter are difficult to understand
because interparticle collisions are strongly inelastic. If a high
density of particles builds up in a region because of random
fluctuations, the collision rate and therefore the rate of energy
loss increases in the region.  This can trap particles in the region,
causing the density fluctuations to grow.~\cite{Goldhirsch} Experimentally, one
observes distinctive phenomena such as force chains and stability
against mechanical collapse in very sparse static packings,
non-Maxwellian velocity distributions~\cite{vanNoije,Rouyer,Reis} and inelastic collapse in
dilute granular gases,~\cite{McNamara,Hopkins} and shear thinning and shear thickening in
the intermediate regime.~\cite{Brown}

The tendency of flowing granular matter to get `jammed' and stop
flowing at low densities is a practical problem that limits the
flow rate in the industrial use of granular materials.~\cite{LiuNagelCRCBook} Remarkably,the
transition from a flowing to a jammed state in granular matter,
structural glasses, and foams and colloids, can be studied with a
unified approach.~\cite{LiuNagel1998} When the transition occurs at zero temperature
and zero shear stress as the density is varied, the transition point
is called ``Point J",~\cite{OHern} and is characterized by diverging length
scales~\cite{Wyart,Ellenbroek} suggestive of a second order phase transition. At the same
time, other properties of the system change discontinuously at Point
J,~\cite{OHern} as one would expect at a first order phase transition.

The density of states for vibrational modes in a granular system
is one of the properties that has a signature of the transition at
Point J. A jammed granular system has mechanical rigidity. Even
though the force between two particles is a nonlinear function of
the compression between them, the small deviations from the jammed
state (which already has non-zero compression) can be analyzed using
a linear model, resulting in normal modes. Extensive numerical
simulations~\cite{OHern} on systems at zero temperature and zero shear stress
show that the density of states $D(\omega)$ as a function of $\omega$
approaches zero linearly as $\omega\rightarrow 0$ if the particle
density is greater than the critical density.
As the particle density is reduced, the
slope of $D(\omega)$ at the origin becomes steeper, until at Point
J, $D(\omega \rightarrow 0) \neq 0.$

In the linearized analyis of vibrational modes, the system can be
treated as a network of random springs, with the number of springs
decreasing as Point J is approached. It is natural to analyze the
problem using random matrices, and see how the resultant density
of states evolves near the transition. This has been done,~\cite{beltukov} and
yields a broad peak in the density of states that reaches $\omega
= 0$ as the transition is approached. However, the model also
predicts a gap in the density of states near $\omega = 0$ above the
transition, which does not match the numerical results. There are a few 
other qualitative discrepancies. 

Although it is encouraging that some features of the random matrix
density of states matches $D(\omega)$ from numerical simulations,
it is well known that the overall density of states predicted by
random matrix theory often differs from what is observed in systems
to which the theory applies, because of non-universal effects \cite{mehta}.
Instead, the correlations in the density of states and the distribution
of level spacings is a more reliable indicator of the validity of
the random matrix approach \cite{mehta}.

In this paper we therefore
turn to the correlations in the density of states predicted by random matrix
theory. We argue that the Laguerre orthogonal ensemble rather than the Gaussian
orthogonal ensemble (GOE) is the appropriate random matrix model for granular
bead packs. The correlation
function for the Laguerre ensemble differs from that for the GOE near the
low-frequency edge of the allowed range of $\omega$ \cite{nagao}. 
By comparing the correlations in the numerically computed vibrational
spectrum of granular bead packs near the jamming transition to the
predictions of the Laguerre ensemble and the GOE we are able to 
demonstrate good agreement with the former and to exclude the latter.

The distribution 
of consecutive level spacings is also a universal feature of the spectrum
that should be described by random matrix theory. We find that the level
spacing distribution predicted by the Laguerre ensemble is very close
to the GOE result both near the zero frequency edge and at high frequency. 
The spectra calculated for granular bead packs in dynamical simulations
are found to agree with this distribution. This finding further validates
the random matrix approach but it does not help discriminate between the 
Laguerre ensemble and the GOE. The agreement of the level spacing distribution 
with the GOE result has been observed earlier,~\cite{Silbert,Nelson} but without 
reference to the Laguerre ensemble.

We also construct a random lattice model, which is a physically
motivated variant of the random matrix ensemble. Although it is not
possible to calculate the properties of this model analytically,
numerical results reveal that all the qualitative features of
$D(\omega)$ are reproduced. At the same time, the correlation
functions and the level spacing distribution seen in the idealized
random matrix theory are not significantly changed.

The rest of the paper is organized as follows. 
In section II we summarize the random matrix approach to the problem
and the results for the density of states. In Section III we compare
the autocorrelation function for the density of states and the level
spacing distribution for the Laguerre ensemble and the GOE to the
vibrational spectra for granular packs near point J, obtained from
dynamical numerical simulations. In Section IV we introduce the
random lattice model and show that it reproduces both universal
and non-universal features of vibrational spectrum of granular
packs. In Appendix A we present an analysis of the level spacing
distribution for the Laguerre ensemble and in Appendix B some 
technical details regarding the autocorrelation.

\section{Laguerre ensembles}
We follow the approach of Ref.~\cite{beltukov} here. Within linear response,
if the particles in the granular assembly are displaced slightly
from their resting positions, their accelerations are of the form
$\ddot x = -K x,$ where $x$ is a $N$ component column vector and $K$ is a 
$N\times N$ matrix. The crucial observation~\cite{lubensky,calladine} is that the connection
between accelerations and displacements is a two-step process.
Within linear response, each contact between a pair of particles
can be represented as a spring that has been precompressed by some
amount. Thus one has a network of springs, with various spring
constants. When a particle is displaced, it stretches (or compresses)
each spring that it is connected to, by an amount that is equal to
the component of its displacement along that spring. The spring
exerts a restoring force that is proportional to this stretching;
the spring constant can be different for each spring. Thus we have
\begin{equation}
	f_i = k_i \tilde A_{ij} x_j
\end{equation}
where $\tilde A_{ij} = \cos\theta_{ij}$ if the $i$'th spring is
connected to the $j$'th particle, with $\theta_{ij}$ being the angle
between the displacement and the direction of the spring, and $\tilde
A_{ij} = 0$ otherwise. The restoring force on each particle is the
sum of the forces from all the springs it is connected to, so that
\begin{equation}
	m_j \ddot x_j = - \tilde A^T_{ji} f_i
\end{equation}
i.e.
\begin{equation}
	\ddot x_j = - \frac{1}{m_j} \tilde A^T_{ji} k_i \tilde A_{ik} x_k.
\end{equation}
Defining $A_{ij} = \sqrt k_i \tilde A_{ij}/\sqrt m_j,$ this is equivalent to 
\begin{equation}
	\ddot x_j = - A^T_{ji} A_{ik} x_k
	\label{AAT}
\end{equation}

Even though we have presented the argument as if the displacement
of each particle is one scalar variable, it is easy to extend this
to particles in a $d$-dimensional system, with $N$ displacement
variables for $N/d$ particles. We have implicitly assumed that the
particles are frictionless spheres, so that torque balance is
trivially satisfied.

The matrix $A$ is a rectangular matrix, since the number of springs
is greater than $N.$ As one approaches Point J, the number of
contact forces decreases, being equal to $N$ at the transition.

In the random matrix approach to this problem, we assume that all
the entries in the matrix $A$ are independent Gaussian random
variables, drawn from a distribution with zero mean and (with a
suitable rescaling) unit variance.  This is the Laguerre random
matrix ensemble. It can be shown~\cite{Forrester} that the reduced probability density
for the eigenvalues $\omega_1^2, \omega_2^2\ldots \omega^2_N$ of
$A^T A$ is of the form
\begin{equation}
	p(\{\omega\}) \propto \prod_{i < j} |\omega_j^2- \omega_i^2|\prod_i \omega_i^{M - N} \exp[-  \sum_i \omega_i^2/2]
	\label{gaussian}
\end{equation}
where the matrix $A$ is a $M\times N$ dimensional matrix, i.e. there
are $M$ inter-particle contacts in the system, and without loss of
generality we have chosen all the $\omega_i$'s to be greater than
zero.

Rewriting the probability density in terms of $\lambda_i =
\omega_i/\sqrt N,$ we have
\begin{equation}
	p(\{\lambda\}) \propto \exp\left[-  N \sum_i \frac{\lambda_i^2}{2} + f \ln |\lambda_i| + \sum_{i < j} \ln |\lambda_i^2 - \lambda_j^2|\right]
\end{equation}
where $M/N = 1 + f.$
If $M - N$ is $O(N),$ a saddle-point expansion yields
\begin{equation}
	 \frac{f}{\lambda} - \lambda + P\int_0^\infty \rho(\lambda^\prime)\left[\frac{1}{\lambda - \lambda^\prime}  + 
	\frac{1}{\lambda + \lambda^\prime}\right]d\lambda^\prime 
\end{equation}
wherever $\rho(\lambda) \neq 0.$ Here $\rho(\lambda)$ is the density of
eigenvalues, normalized to $\int_0^\infty \rho(\lambda) d\lambda =
1,$ and the $P$ denotes the principal value of the integral.
Symmetrizing $\rho(\lambda)$ by defining $\rho(\lambda < 0) =
\rho(-\lambda),$ the function
\begin{equation}
	F(\lambda) =	\int_{-\infty}^\infty \frac{\rho(\lambda^\prime)}{\lambda - \lambda^\prime} d\lambda^\prime
\end{equation}
of the complex variable $\lambda$ is analytic everywhere except
that it has branch cuts on the real line over intervals where
$\rho(\lambda)\neq 0,$ where it is equal to
\begin{equation}
\lambda - \frac{f}{\lambda} \mp i \pi \rho(\lambda).
\end{equation}
Furthermore, $F(\lambda\rightarrow 0)$ is finite, and
$F(\lambda\rightarrow\infty) \rightarrow 2/\lambda,$ because the
symmetrized extension of $\rho(\lambda)$ integrates to 2. This has
the solution
\begin{equation}
	F(\lambda) = \lambda - \frac{f}{\lambda} - \frac{\sqrt{(\lambda^2 - a^2) (\lambda^2 - b^2)}}{\lambda}
\end{equation}
with 
\begin{eqnarray}
	a &=& \sqrt{M/N} - 1\nonumber\\
	b &=& \sqrt{M/N} + 1.
	\label{gap}
\end{eqnarray}
The density of states is then
\begin{equation}
	\rho(\lambda) = \frac{1}{\pi} \frac{\sqrt{(b^2 - \lambda^2)(\lambda^2 - a^2)}}{\lambda} \qquad a < \lambda < b
\end{equation}
where we have removed the extension of $\rho(\lambda)$ to $\lambda
< 0,$ so that $\int_0^\infty \rho(\lambda) d\lambda = 1.$ The density
of states $D(\omega)$ has the same form, but with $a$ and $b$
rescaled, which is not significant since Eq.(\ref{gaussian}) was
already obtained after rescaling.

When $M/N > 1,$ there is a broad peak in $D(\omega),$ with a gap
in the spectrum near $\omega = 0.$ The peak is not symmetric, falling
off much more sharply on the small $\omega$ side than on the large
$\omega$ side. In the middle, the peak slopes downwards as $\omega$ is increased.
As $M/N$ is reduced, the gap shrinks while the width
of the peak remains constant. When $M/N = 1,$ $D(\omega) =
\sqrt{\Omega^2 - \omega^2}/\pi$ which matches the Wigner semicircle
law for the Gaussian orthogonal ensemble, and $D(\omega=0)\neq 0.$

One can compare these analytical predictions with numerical results.
Simulations of two-dimensional frictionless soft spheres are
performed, allowing the spheres to equilibrate from an initial
random configuration~\cite{Corey}. As with the analytical prediction,
there is a broad peak, that falls off more sharply at small $\omega$
than at large $\omega.$ The density of states $D(\omega=0) = 0$
except at the transition. However, the numerical data does not show
the gap in the spectrum near $\omega = 0$ predicted by random matrix
theory. The numerical data also has a pronounced boson peak at a non-zero
value of $\omega,$ and a cusp in $D(\omega)$ at the origin at the
transition. None of these is consistent with the prediction from\
random matrix theory. We return to this point in Section~\ref{sec:lattice}.

\section{Correlations}

In this section we subject the predictions of the random matrix
model to more stringent and appropriate tests. We have seen in the previous
section the mean density of states of the random matrix model
does not exactly match the density of states of the dynamical
numerical simulation of a granular pack. However that is not the
appropriate test of a random matrix model. In all of its
successful applications, what random matrix theory is able to
predict correctly is not the mean density of states, but
rather statistical features like the density of states
correlation and the level spacing distribution, that are
computed after the spectrum has been smoothed by a procedure
called unfolding (explained below) that makes the mean density
of states uniform.

Theoretically one can understand this as follows.
It can be shown~\cite{brezin} that, for an ensemble of random matrices, the mean
density of states can be changed at will by varying the assumed 
distribution of the matrix elements, but that the correlations of the
unfolded spectrum retain a universal form that depends only on the
symmetries of the ensemble considered (such as whether the matrices
are real or complex or quaternion real). From this point of view
the mean density of states we obtained in the previous section is 
simply an artifact of the Gaussian distribution we chose for our
random matrix elements, but the unfolded density of states correlations
and the level spacing distribution are universal and should match
the numerical data for jammed granular matter if the model is applicable.

The key feature of the random matrix spectrum is that it is rigid
(i.e. highly correlated). The rigidity of the spectrum is revealed at
small energy scales by the distribution of consecutive level spacings.
The longer range rigidity can be demonstrated by the autocorrelation 
of the density of states or by specific statistical measures such as the
number statistic and the spectral rigidity~\cite{mehta}.

Insight into the strong correlations between the eigenvalues 
implied by the Laguerre ensemble 
distribution Eq.(\ref{gaussian}) is provided by
the following plasma analogy. We focus on the case $M = N+1$ since
we are interested in the spectrum for point J. If we rewrite
rewrite the factor $| \omega_j^2 - \omega_i^2 |$ in Eq.(\ref{gaussian})
as $\exp(\ln |\omega_j - \omega_i | + \ln |\omega_j + \omega_i |)$
and the factor $\omega_i$ as $\exp (\ln \omega_i )$, 
we can interpret Eq. (\ref{gaussian}) as the partition function
of a classical plasma of N particles located on the positive
$\omega$ axis at the locations $\omega_1, \ldots, \omega_N$ with 
logarithmic interactions between the particles as well as
logarithmic interactions between each particle and image
particles at locations $-\omega_1, \ldots, -\omega_N$. In
the plasma analogy the particles are also confined near the
origin by a quadratic potential and are constrained to remain 
on the positive $\omega$ axis by a hard wall at the origin. 

\begin{figure}
\begin{center}
\includegraphics[width=\columnwidth]{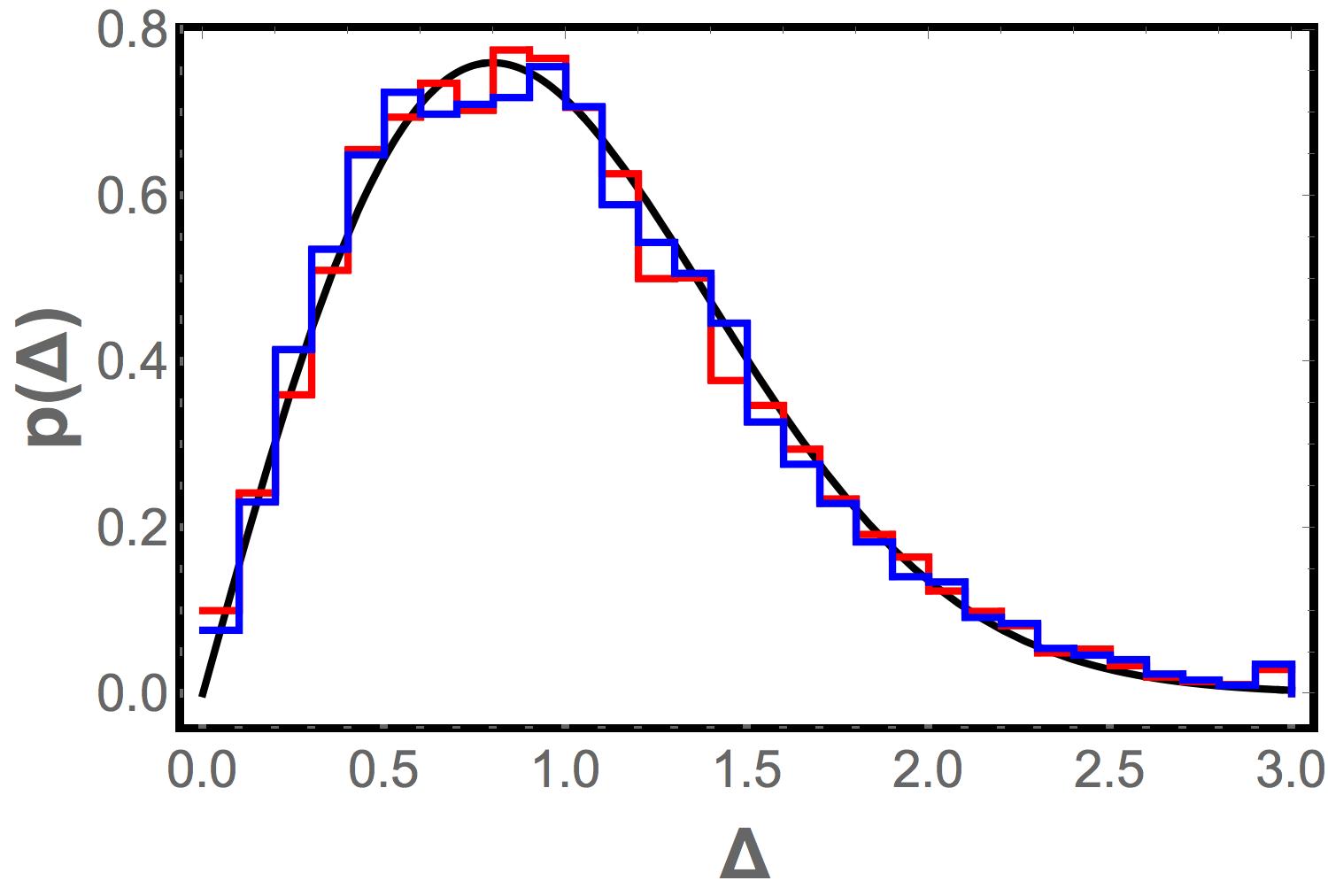}
\end{center}
\caption{
The red and blue histograms show the consecutive level
spacing distribution for the numerically computed vibrational spectra
of jammed granular material. An ensemble of one thousand realizations
of the jammed state was used. The red histogram bins the eleven
consecutive level spacings between the frequencies $\omega_5$ through
$\omega_{16}$ for each realization; the blue histogram eleven
consecutive spacings between frequencies $\omega_{400}$ through
$\omega_{411}$. Each spacing $\omega_{i+1} - \omega_i$ is normalized
by $\langle \omega_{i+1} - \omega_i\rangle,$ where the average is
taken over the one thousand realizations. The black curve corresponds to
the Wigner surmise for the level spacing distribution of the Gaussian
orthogonal ensemble (which is indistinguishable from the Laguerre
ensemble at this level of resolution).  The close agreement between
the two histograms and the solid black curve are consistent with
the predictions of our random matrix model of the jammed state of
granular matter.}
\label{fig:spacing} 
\end{figure}

We turn now to the distribution of spacings between consecutive
levels. Intuitively one might expect that the spacing between
consecutive levels at low frequency would be different from
that between two levels at high frequency. This is because,
in terms of the plasma analogy, in the first instance the two
interacting levels are near the hard wall, whereas in the latter
instance they are deep in the interior of the plasma. However,
we show in Appendix A that the consecutive level spacing
distribution is not noticeably different for high and
low frequencies, and neither of these is noticeably different 
from the distribution for the GOE.

In Fig. \ref{fig:spacing} we compare the predictions of the
Laguerre model to the numerically computed jammed granular
spectrum. The red histogram bins the eleven consecutive 
level spacings between the frequencies $\omega_5$ through $\omega_{16}$
for each realization; each spacing is normalized by $\langle \omega_{i+1} - \omega_{i} \rangle,$
where the average is taken over the one thousand realizations of the
jammed granular pack. The blue histogram is the same but for the
eleven level spacings between the frequencies $\omega_{400}$ through $\omega_{411}$.
The two distributions are seen to be indistinguishable and to be in
good agreement with the approximate analytic formula for the Laguerre
ensemble (solid black curve) which is itself indistinguishable from
the prediction of the GOE at the resolution of the figure. 

Fig.~\ref{fig:spacing} shows that the numerical level spacing distribution
is consistent with the prediction of the Laguerre random matrix
model and hence is a validation of that model. However, the level
spacing distribution is not able to distinguish between the Laguerre
model and the GOE, and is therefore a less sharp test of the Laguerre
model than the density of states autocorrelation to which we now
turn. 

In order to calculate these correlations it is convenient to rewrite
the distribution in Eq. (\ref{gaussian}) for the case $M = N + 1$ in
the form
\begin{equation}
    P(x_1, x_2, \ldots, x_N) \propto \prod_{i \neq j} | x_i - x_j | \prod_{k} \exp( - x_k)
    \label{eq:laguerre}
\end{equation}
where $x_i = \omega_i^2$. The one-point and two-point correlation
functions are defined as
\begin{equation}
    R_1 (x) = N \int_0^\infty d x_2 \ldots \int_0^\infty d x_N P(x, x_2, \ldots, x_N)
    \label{eq:onepoint}
\end{equation}
and 
\begin{equation}
    R_2 (x, y) = N(N-1) \int_0^\infty d x_3 \ldots \int_0^\infty d x_N P(x, y, x_3,\ldots, x_N)
    \label{eq:twopoint}
\end{equation}
The plasma analogy shows that calculation of the correlation
functions in Eq. (\ref{eq:onepoint}) and Eq. (\ref{eq:twopoint}) is a
formidable problem in classical statistical mechanics.  Nonetheless
it has been exactly done by Nagao and Slevin~\cite{nagao} by rewriting Eq. (\ref{eq:laguerre}) in the
form of a quaternion determinant and performing the integrals by a
generalization of a theorem of Dyson~\cite{dyson} on integration over quaternion
determinants. Before we give those results we first describe the
unfolding procedure.

$R_1(x)$ is evidently the density of states, and we now define
\begin{equation}
    \xi (x) = \int_0^x d x^\prime \; R_1 (x^\prime)
    \label{eq:staircase}
\end{equation}
where $\xi(x)$ is the cumulative density of states. 
The unfolded two point correlation function is then defined as
\begin{equation}
	L_2 (\xi_1, \xi_2) = \frac{1}{R_1[ x (\xi_1) ]} \frac{1}{R_1 [ x(\xi_2) ]} R_2 [ x (\xi_1), x(\xi_2)]
    \label{eq:ell2}
\end{equation}
It is easy to see that if $L_1 (\xi)$ is similarly defined then
$L_1 (\xi ) = 1$ showing that after unfolding the density of states
is uniform with a mean level spacing of unity. The exact expression
for $L_2$ is rather lengthy and is relegated to Appendix B. 

\begin{figure}
	\begin{center}
		\includegraphics[width=\columnwidth]{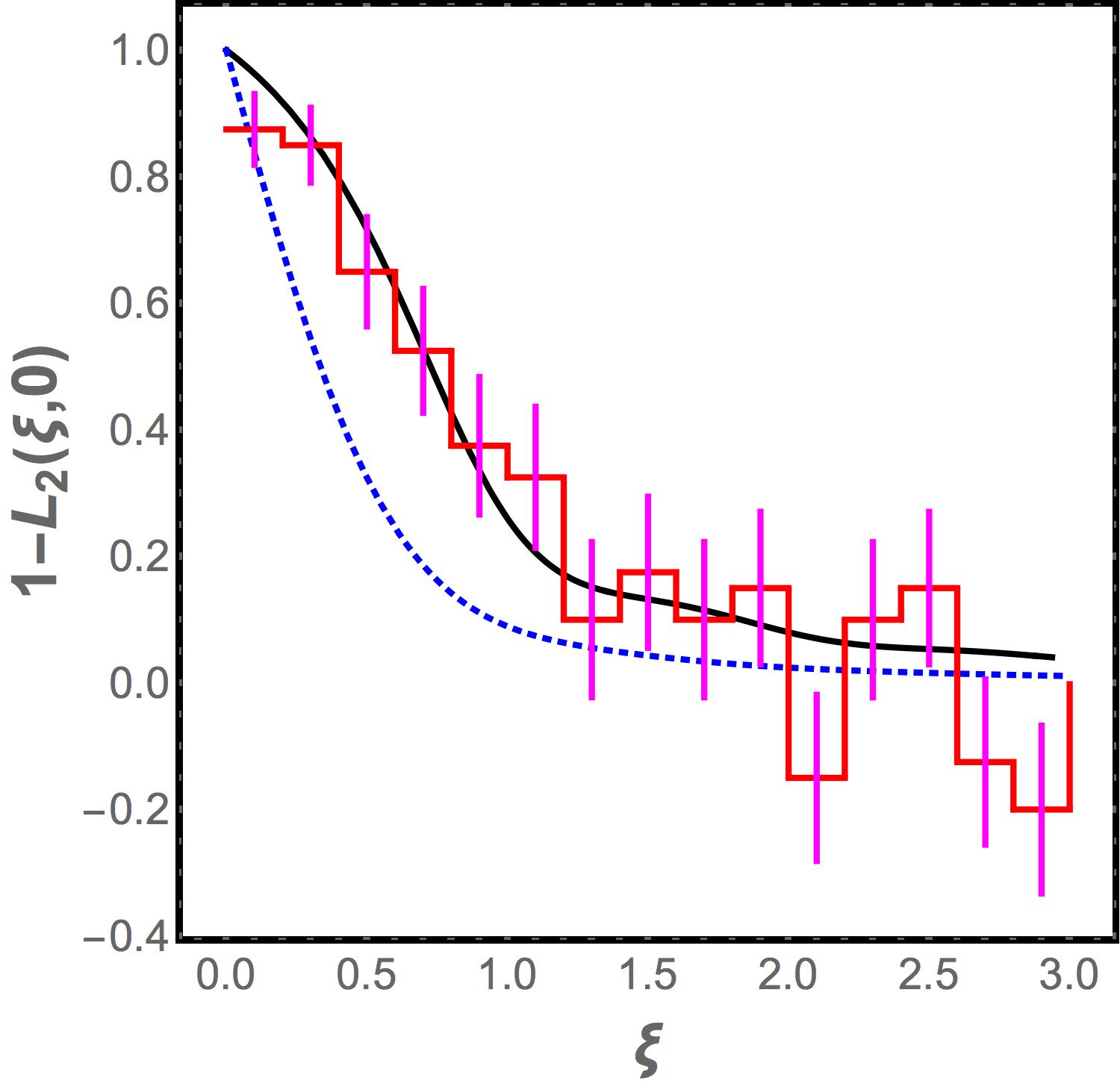}
	\end{center}
	\caption{Correlation function $1 - L_2(\xi, 0)$ from random matrix theory as a function of $\xi,$ out 
	to $\xi=3.0,$ i.e. out to a distance within which there should, on average, be three
	eigenvalues. This is compared to a histogram of the vibrational frequencies from the numerical 
	simulations. The analytical results for the Gaussian orthogonal ensemble and the Laugerre ensemble 
	are both shown, and the latter is clearly superior.} 
	\label{fig:corrln}
\end{figure}

In Figure~\ref{fig:corrln} we plot $1 - L_2(\xi, 0)$ as a function
of $\xi$ for the Laugerre ensemble. The corresponding plot for the
Gaussian Orthogonal ensemble is also shown. When $\xi$ is large,
the two curves approach each other.  Indeed, the analytical expression
for $1 - L_2(\xi)$ for $\xi \gg 1$ in the Laguerre ensemble can be
verified to be
\begin{eqnarray}
1 - L_2 (\xi, 0) & = & \frac{\sin^2 \pi \xi}{ ( \pi \xi )^2} \nonumber \\
& + & \left[ \frac{1}{2} - \int_0^\xi d y \frac{ \sin \pi y}{\pi y} \right] 
\left[ \frac{ \cos \pi \xi}{\xi} - \frac{1}{\pi \xi^2} \sin \pi \xi \right]
\nonumber \\
\label{eq:goetwopoint}
\end{eqnarray}
which coincides with the form for the same quantity in the GOE.
Intuitively, the reason for this coincidence can be understood in
terms of the plasma analogy. If the particles are deep in the
interior of the plasma the edge effects produced by
the image charges are screened and the Laguerre plasma becomes
indistinguishable from the GOE plasma.

Although the two curves coincide in the asymptotic limit
$\xi \gg 1$, Fig~\ref{fig:corrln} shows that there is a range
of $\xi$ values where the predictions of the Laguerre ensemble
differ significantly from the GOE. Hence comparison to the 
correlation function for the numerical data for the jammed
granular spectrum provides a stringent test that is able
to distinguish between the Laguerre ensemble and the GOE.

The numerical data are analyzed as follows. 
The vibrational frequencies obtained in the numerical simulations
from all the 1000 realizations of the jammed state are merged
together, and bins are constructed with 200 eigenvalues in each,
i.e. there is an average of 0.2 eigenvalues per realization of the
jammed state in each bin. Next, we calculate
\begin{equation}
	1 - \frac{1}{(0.2)^2} [\langle n_0 n_i\rangle - \langle n_0 \rangle\delta_{i0}]
\end{equation}
where $n_i$ is the number of vibrational frequencies in the $i^{{\rm th}}$ bin
in any given realization, and the average is over the realizations
of the jammed state. The histogram of the values obtained for this
discretized correlation function are compared with the analytical
prediction from the Laugerre ensemble and the GOE, and as seen in Fig.~\ref{fig:corrln}, the Laguerre
ensemble fits the data very well within the error bars (while the
GOE does not).

\section{Random Lattice Model}
\label{sec:lattice}
As discussed earlier, the extent to which the density $D(\omega)$
of vibrational frequencies for jammed granular materials agrees
with the predictions of random matrix theory is not a good test of
the applicability of random matrix theory to these materials, because
the distribution of eigenvalues is a non-universal prediction of
random matrix theory: if the random matrices are not assumed to be
Gaussian, the density of eigenvalues changes.

Nevertheless, there are qualitative discrepancies between the
numerically measured $D(\omega)$ and the density of eigenvalues
$\{\omega_i\}$ obtained from random matrix theory, that are worth
trying to address. As seen in Eq.(\ref{gap}), there is a gap in the
spectrum of eigenvalues near $\omega = 0$ above the jamming transition,
where $M > N$ in Eq.(\ref{gap}).  By contrast, in the numerical simulations,
$D(\omega)$ is only zero at $\omega
= 0$ (except at the jamming transition), and increases linearly for
small $\omega.$  Second, at the jamming transition, $D(\omega)$ has
a cusp-like peak at $\omega = 0,$ while random matrix theory predicts
a flat $D(\omega)$ near $\omega = 0$ at the transition.  Finally, the numerical
$D(\omega)$ has a pronounced boson peak at $\omega = \omega_0\neq 0,$
which is not reproduced by random matrix theory.

Various attempts have been made to construct random matrix models
that reproduce the properties of granular systems more closely.
Ref.~\cite{Manning} studies several different random matrix ensembles
and their effect on the structure of eigenmodes with frequencies
in the boson peak. Ref.~\cite{Middleton} uses weighted Laplacian
dynamical matrices to reproduce an intermediate regime in $D(\omega)$
(between the boson peak and the low frequency behavior) and $\sim
\omega^4$ scaling of the density of states in this regime.
Ref.~\cite{Beltukov1} uses a combination of a random and a regular
matrix for the dynamical matrix, to eliminate the gap near $\omega
= 0.$ Also, Ref.~\cite{Parisi} has studied an abstract model that
they argue is in the appropriate universality class.

In Eq.(\ref{AAT}), we have assumed that the entries in the matrix
$A$ are all independent random variables drawn from a Gaussian
distribution. In reality, since the matrix $A$ is supposed to be a
mapping from coordinates to contact forces, and only two particles
are associated with a contact, only the entries associated with two
particles (with $d$ entries per particle for $d$-dimensional particles)
should be non-zero in any column of $A$. Thus $A$ should be a sparse
matrix.

One could choose the two particles associated with each force
randomly, but this would result in the system breaking up into separate
clusters, not connected to each other, leading to an overabundance of zero
modes. Moreover, the concept of adjacency would not be respected: two randomly
chosen particles would be likely to be far apart, and should not have been
allowed to share a contact. 

Instead of choosing the particles associated with a force randomly,
we approximate the system as being equivalent to a
triangular lattice (with periodic boundary conditions), but with
each particle displaced from the position where it would be in a perfect
triangular lattice. This randomizes the orientation of the contacts between 
particles. 

To be specific, particles are arranged in successive horizontal
layers, with each particle having contacts with the two particles
immediately below it: slightly to the left and slightly to the
right. Shifting the numbering in each row by half a lattice spacing relative to its
predecessor, the particle $(i, j)$ connects to the particles numbered $(i, j - 1)$ and $(i
+ 1, j - 1)$ with periodic boundary conditions in both directions.
(A particle in the bottom layer, $(i, 1),$ connects with $(i - L/2, L)$ and $(i + 1 - L/2, L)$ in 
the topmost layer, where $L$ is the number of layers.)
In addition, each particle has a probability of connecting to its
neighbor on the right in the same row: $(i, j) \rightarrow (i + 1,
j).$ All contacts are bidirectional, i.e.  each particle is connected
to two particles in the row above it, two in the row below it, and
either zero, one or two adjacent particles in the same row. The
spring constant associated with each contact is chosen randomly,
and the bond angles are also chosen randomly with the constraint
that the bond connecting a particle to its left (right) neighbor
in the row below points down and to the left (right), while the bond
connecting a particle to its neighbor in the same row on the left (right)
is more horizontal than the bond to its neightbor below and to the
left (right). This model is similar to the model introduced for free-standing
granular piles~\cite{Narayan}, a vector generalization of the scalar-force
``q-Model" used to model such systems~\cite{Coppersmith}.

This Random Lattice Model (RLM) with $24 \times 24$ sites was simulated in this manner, and
the vibrational frequencies from 100 different realizations of
randomness were merged and plotted as a histogram. In two-dimensions,
the number of coordinate degrees of freedom is $N = 2 \times 24\times
24 = 1152,$ which is comparable to the 800 frequencies that were present
in each system in the dynamical simulations~\cite{Corey}. The ratio
of the number of contact forces to the number of coordinate degrees
of freedom, which corresponds to $M/N,$ increases from 1 to 1.5 as
the probability of establishing contacts within the same layer
increases from 0 to 1.

\begin{figure}
	\begin{center}
		\includegraphics[width=\columnwidth]{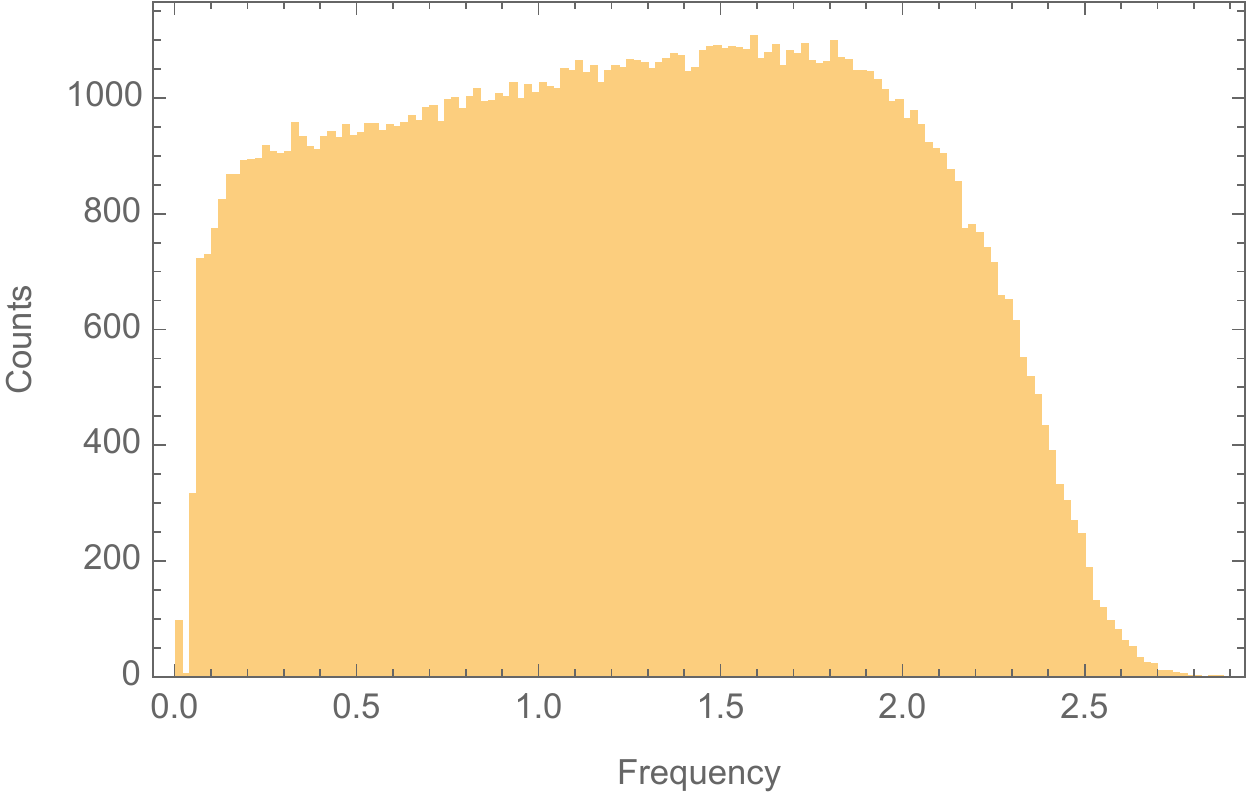}
		\includegraphics[width=\columnwidth]{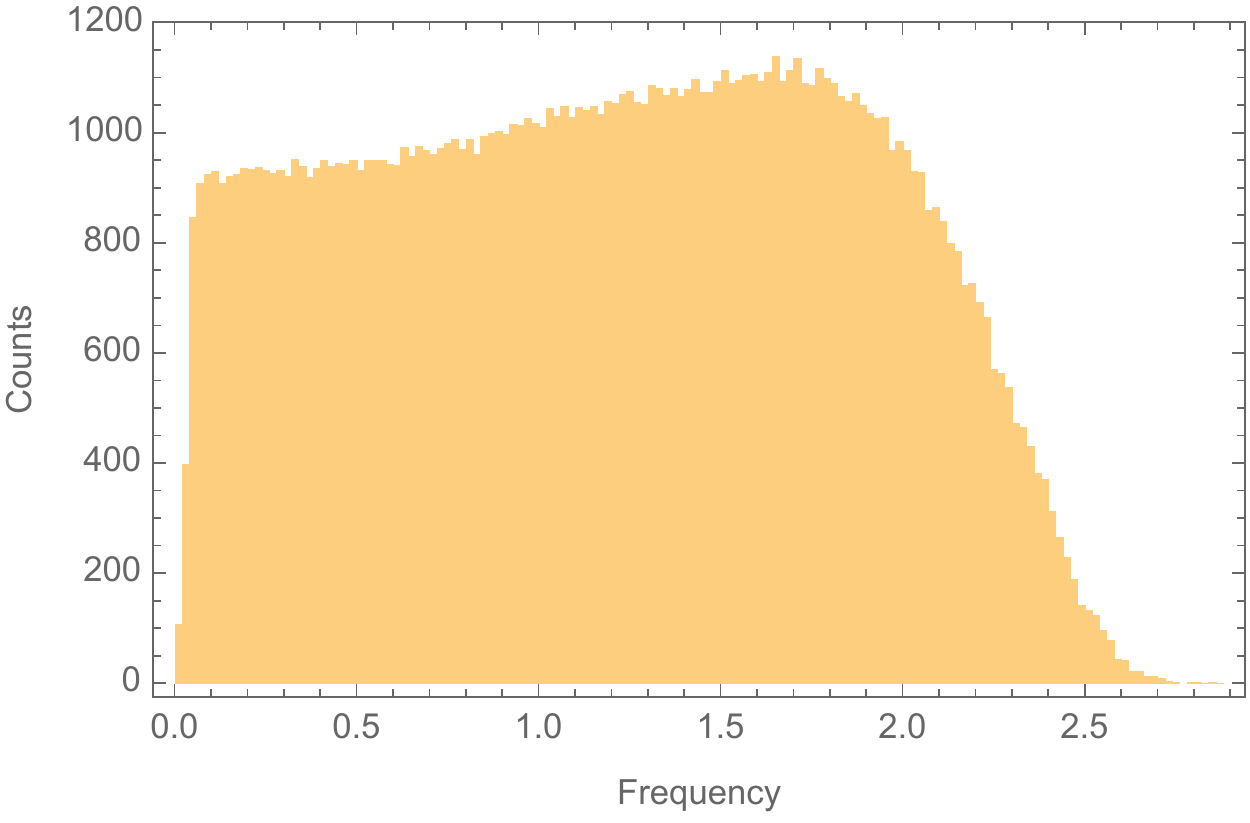}
		\includegraphics[width=\columnwidth]{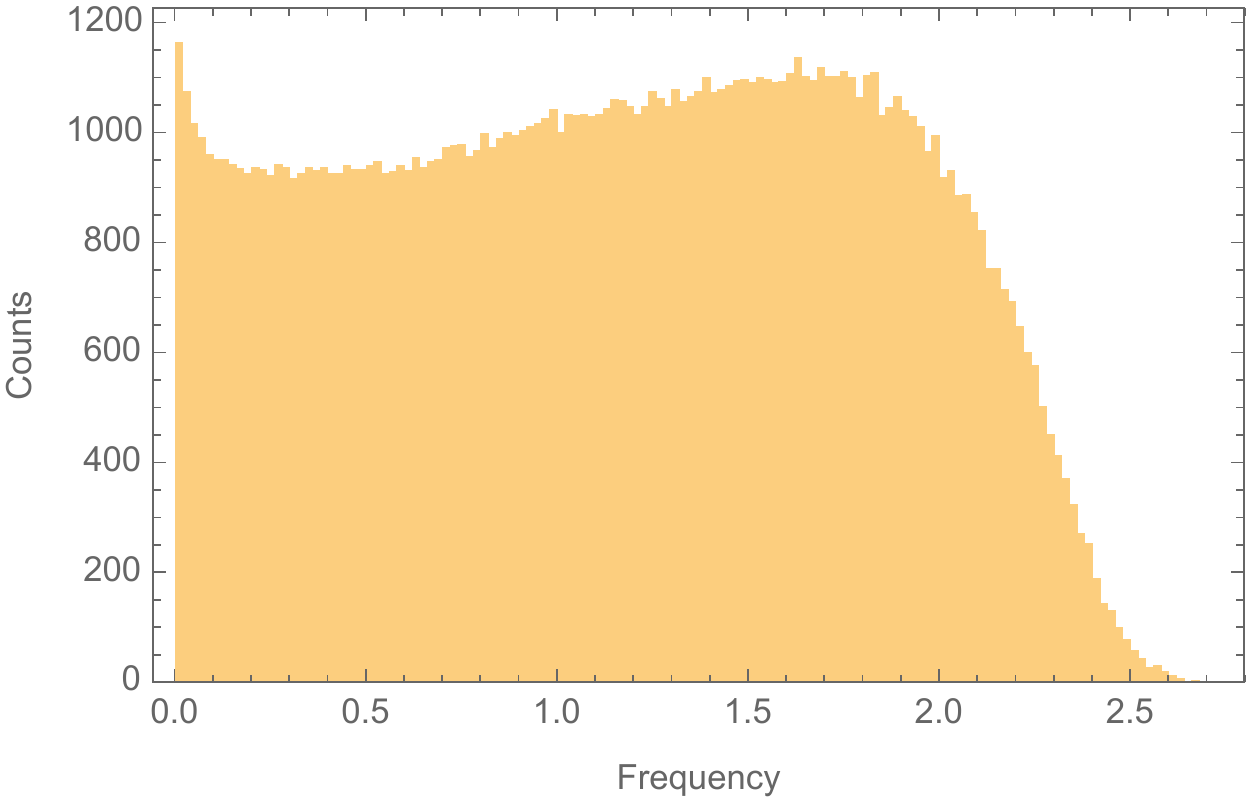}
	\end{center}
	\caption{Histogram of vibrational frequencies obtained from the random 
	lattice model described in this paper. 100 realizations of 24 x 24 random
	triangular lattices were constructed, with the ratio $M/N$ equal to 
	1.1 (top), 1.05 (middle) and 1.0 (bottom) respectively.}
	\label{fig:randomlattice}
\end{figure}
The results are shown in Figure~\ref{fig:randomlattice}.  The
transition from $D(\omega = 0) = 0$ when $M/N \neq 1$ to $D(\omega = 0) \neq 0$ 
when $M/N = 1,$ which is 
seen in random matrix theory and in the dynamical simulations, is also seen in 
the RLM density of states. But in addition, 
there is a cusp in $D(\omega=0)$ at the transition point, as is
seen in the dynamical simulations~\cite{Corey}.  The boson peak
at $\omega\neq 0$ seen in the simulations is also reproduced for
the lattice model, being most pronounced at the transition point.

Although there is still a gap in the Random Lattice Model frequency
spectrum near $\omega = 0,$ it is approximately half what one would
predict from random matrix theory for the corresponding $M/N.$
Moreover, it is clear that it is a finite size effect: the global
translational invariance of the random lattice with periodic boundary
conditions ensures that there are two zero modes (seen clearly in
the first plot in Figure~\ref{fig:randomlattice}), and the locality
of the connections that are made ensures that long wavelength
oscillations have low frequencies. This is verified by increasing
$N$ and checking that the gap decreases even though $M/N$ is held
constant.

\begin{figure}
	\begin{center}
		\includegraphics[width=\columnwidth]{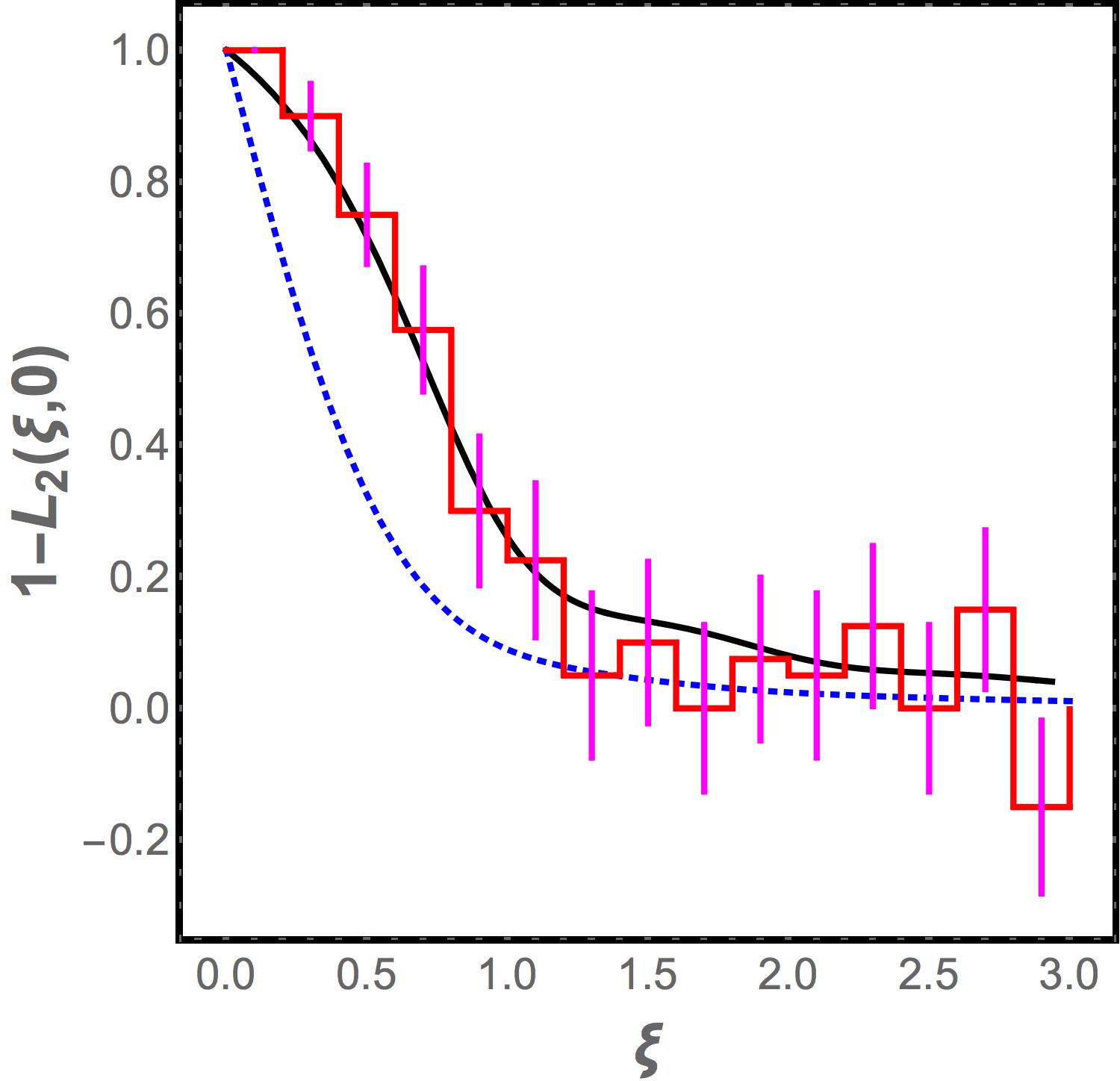}
	\end{center}
	\caption{Correlation function $1 - L_2(\xi, 0)$ for the Random 
	Lattice Model introduced in this paper and for the Laguerre ensemble,
	at the transition point (i.e. with square random matrices). Good
	agreement is seen between the two. The correlation function for the
	Gaussian Orthogonal ensemble is also shown for comparison.}
	\label{fig:RLMcorrlns}
\end{figure}
\begin{figure}
	\begin{center}
		\includegraphics[width=\columnwidth]{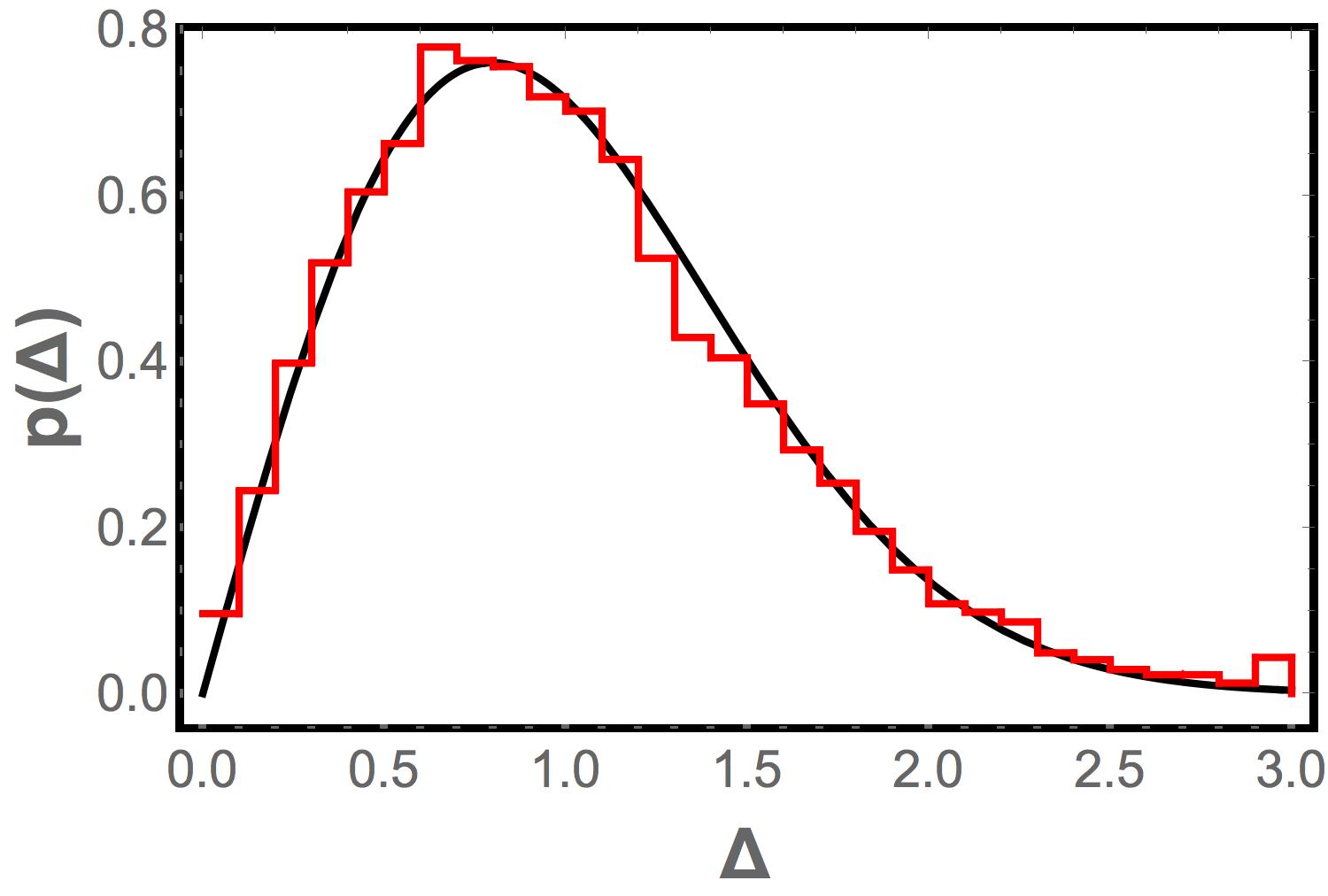}
	\end{center}
	\caption{Level spacings for the Random Lattice Model and a
	fit to the Wigner surmise for the Gaussian Orthogonal
	Ensemble.  Spacings from the fifth to the fifteenth normal
	mode frequencies are normalized, as discussed in the paper,
	and combined to create the histogram. The Wigner surmise
	fits the distribution very well, but as discussed in the
	paper, the fit applies equally to the Laguerre ensemble.}
	\label{fig:RLMspacings}
\end{figure}

We see that the Random Lattice Model has the same change in $D(\omega
= 0)$ at the jamming transition that is seen in the dynamical
numerical simulations and in random matrix theory. However, it also
reproduces other qualitative features of the numerical $D(\omega)$
that random matrix theory did not. As seen in Figure~\ref{fig:RLMcorrlns}
and in Figure~\ref{fig:RLMspacings}, the distribution of spacings
between consecutive frequencies is found to be the same as for the
random matrix ensemble, consistent with the Wigner surmise, and the
correlation function $1 - L_2(\xi,0)$ at $M/N = 1$ matches that
obtained for the Laguerre ensemble. Thus the random lattice model
retains the positive features of random matrix theory, while curing
its problems.

\section{Conclusions}
In this paper, we show that a random matrix approach can be used
successfully to calculate the correlations between vibrational
frequencies in a granular system near the jamming transition, if
the matrix ensemble is chosen correctly. By modifying the random
matrices according to physical considerations, a Random Lattice
Model is constructed, which retains the correlation functions of
random matrix theory and also successfully reproduces all the
qualitative features in the density of vibrational frequencies.
Such random lattice models may be more broadly applicable to granular
materials. 

\begin{acknowledgments}
	The authors gratefully acknowledge data provided
	by Kyle VanderWerf and Corey O'Hern for numerical simulations
	on granular systems, to which the models in this paper were
	compared. Useful conversations with Satya Majumdar are also
	acknowledged.
\end{acknowledgments}

\appendix

\section{Level spacing distribution}

\label{sec:lsd}

\begin{figure}[ht]
\begin{center}
\includegraphics[width=3.in]{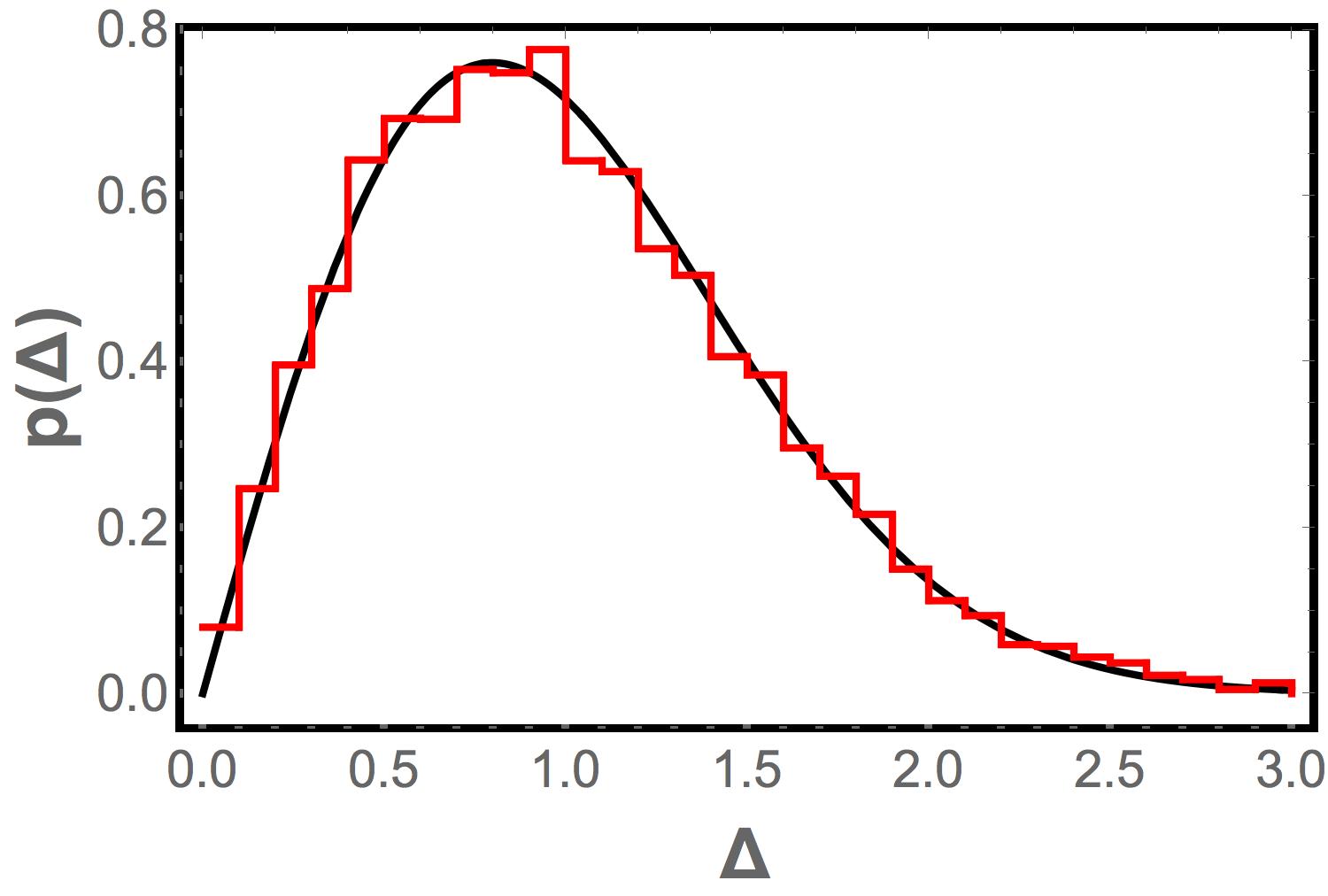}
\end{center}
\caption{Plot of the Wigner surmise for the consecutive level spacing distribution for the Laguerre ensemble
(solid curve). At the resolution of this plot the change in the distribution as a function of
the distance of the levels from the zero frequency edge as well as the deviation from the Gaussian
orthogonal ensemble are invisible. The specific curve plotted in the figure corresponds to $p(\Delta, \omega)$ 
with $\omega = 5.0$; see the discussion in Appendix \ref{sec:lsd}. This curve is seen to be in excellent agreement with a histogram
obtained from a numerical simulation of the Laguerre ensemble.}
\label{fig:surmise}
\end{figure}

In this Appendix we argue that the distribution of consecutive level spacings for the Laguerre ensemble 
Eq. (\ref{gaussian}) is indistinguishable from the GOE. 
To this end it is useful to recall that for the GOE,
Wigner showed that a very good approximation to the level spacing distribution can
be obtained by considering a model with just two levels. This formula, known as the Wigner surmise, is indistinguishable
from the exact result, except in the tails of the distribution where, in any case, the weight is 
negligible, making the distinction irrelevant for applications. Intuitively Wigner's surmise works
because at small spacing the distribution is dominated by the interaction between the two consecutive
levels; the other levels that are neglected in the analysis only matter at large spacings. 
In the same spirit we consider the distribution Eq. (\ref{gaussian}) for the case $M = N + 1 = 3$. 
We define $\Delta = \omega_1 - \omega_2$ as the spacing and $\omega = \frac{1}{2} (\omega_1 + \omega_2)$
as the mean energy of the two levels, and we take $\omega_1 > \omega_2$. For a fixed $\omega$ the
level spacing distribution is then given by 
\begin{equation}
    p ( \Delta, \omega ) \propto \left[\omega^2 \Delta - \frac{1}{4} \Delta^3 \right] \exp \left[- \frac{1}{2} \Delta^2 \right].
    \label{eq:pdeltaomega}
\end{equation}
for $0 < \Delta < 2 \omega$ and $p = 0$ for $\Delta > 2 \omega$. Imposing the condition that the distribution $p$ is normalized
and rescaling the level spacing so that $\langle \Delta \rangle = 1$ we obtain the analog of the Wigner surmise for the
ensemble in Eq. (\ref{gaussian}). Because all the integrals can be worked out in closed form, an explicit but extremely
lengthy formula can be given for $p(\Delta, \omega)$ with appropriate normalization and scaling. 
For the sake of brevity we omit this formula but show in Fig.\ref{fig:surmise} a plot of $p(\Delta, \omega)$ 
for fixed $\omega = 5$. The distribution is essentially indistinguishable from the Wigner surmise for the Gaussian
orthogonal ensemble. The deviation $| p(\Delta, 5.0) - p_{{\rm goe}} (\Delta) | < 2 \times 10^{-4}$ over the 
range of the plot and hence invisible to the eye at the resolution of the figure. 
$\omega$ is a measure of how close the two consecutive levels are to the edge; to be precise
$\int_0^\omega d \Omega R_1 (\Omega)$ is the ordinal number of the pair whose spacing distribution is being 
considered; here $R_1$ is the mean density of states corresponding to the distribution
in Eq. (\ref{gaussian}). As one might expect the distribution approaches that for the Gaussian orthogonal
ensemble as $\omega$ increases. More surprisingly and perhaps disappointingly we find that for pairs of levels
that are quite near to zero frequency also the Gaussian orthogonal ensemble is a good approximation.
This means that in testing whether the spectrum of jammed granular matter is described by 
random matrix theory we cannot use the level spacing to distinguish between our model and the
Gaussian orthogonal ensemble. 

Since the derivation of the level spacing distribution above is non-rigorous we have tested it by
directly simulating our random matrix model. To this end we generated an ensemble of ten thousand
$M \times N$ random matrices $A$ with $M = N + 1 = 101$. The matrix elements are drawn from a Gaussian
distribution with zero mean and unit variance. 
We then evaluated the eigenvalues, $\omega_i^2$ of $A^T A$. Fig.~\ref{fig:surmise} shows a histogram
of the spacing between the fifth and sixth levels, $\omega_6 - \omega_5$, in the different realizations. 
Those spacings have been scaled to have unit mean. As can be seen from the figure the level spacing distribution derived
here as well as the distribution for the Gaussian orthogonal ensemble provide an excellent fit to
the numerical data. Using the numerical data we were also able to test that the distribution for
higher levels is practically indistinguishable, confirming all the features of the analysis above.

It is possible to derive an exact expression for the level spacing distribution using the technology
of quaternion determinants developed by Dyson~\cite{dyson} and Mehta~\cite{mehta} and generalized to the Laguerre ensemble by
Nagao and Slevin~\cite{nagao}. This analysis would allow a more careful study of the tails of the distribution
where it might deviate from the simple approximation derived here. Such an analysis is not needed
for the application considered in this paper but is of intrinsic mathematical interest and we will
return to it elsewhere. 

\section{Correlations of the Laguerre Ensemble}

\label{sec:lagcorr}

Nagao and Slevin~\cite{nagao} have shown that the correlations for the Laguerre ensemble Eq. (\ref{eq:laguerre})
can be computed by rewriting the distribution as a quaternion determinant and performing
the required integrals using a powerful generalization of a theorem by Dyson~\cite{dyson}. 
Here we summarize their results, taking the opportunity to correct some typos in their
paper, and to present the results in a form that does not require the reader to have
familiarity with the specialized language of quaternion determinants or Pfaffians. 

We start by defining the unfolding function 
\begin{equation}
    \xi (x) = \frac{1}{2} \int_0^x d t \; \left[ t J_1^2 (t) - t J_0 (t) J_2 (t) + J_0(t) J_1 (t) 
    \right].
    \label{eq:fone}
\end{equation}
Note that the order of the Bessel function in the first term of the integrand above is given 
incorrectly in Ref.~\cite{nagao}.

Next we define 
\begin{equation}
    r(x) = \sqrt{ J_1^2 (x) + \frac{1}{2} J_0^2 (x) - \frac{1}{2} J_0 (x) J_2 (x) }
    \label{eq:rx}
\end{equation}
and
\begin{equation}
    {\cal C} (x, x^\prime) = 2 \frac{x J_1 (x) J_0 (x^\prime) - x^\prime J_0 (x) J_1 (x^\prime) }{(x-x^\prime)(x + x^\prime)} - 
    \frac{J_0(x) J_1(x^\prime)}{x^\prime}.
    \label{eq:cals}
\end{equation}
 
In terms of these functions one can write down
\begin{equation}
    S (\xi, \xi^\prime ) = \frac{1}{r(x) r(x^\prime)} {\cal C} (x, x^\prime)
    \label{eq:ess}
\end{equation}
where it is understood that $x$ is short for $x(\xi)$ and $x^\prime$
for $x (\xi^\prime)$, the inverse of the function given by Eq.~(\ref{eq:fone}).
We also write
\begin{equation}
    I (\xi, \xi^\prime) = \frac{1}{r(x) r(x^\prime)} \left[ \int_{x^\prime}^x d t \; \frac{t}{2} {\cal C}(x, t) - 
    \theta (x - x^\prime) + \frac{1}{2} 
    \right]
    \label{eq:eye}
\end{equation}
where $\theta$ denotes the unit step function and 
\begin{equation}
    D (\xi, \xi^\prime) = \frac{2}{ x r(x) r(x^\prime)} \frac{\partial}{\partial x} {\cal C}(x, x^\prime);
    \label{eq:dee}
\end{equation}
the variable of integration and the arguments of the integrand in Eq.~(\ref{eq:eye}) are 
given incorrectly in Ref.\cite{nagao}. 

\begin{figure}[ht]
\begin{center}
\includegraphics[width=3.in]{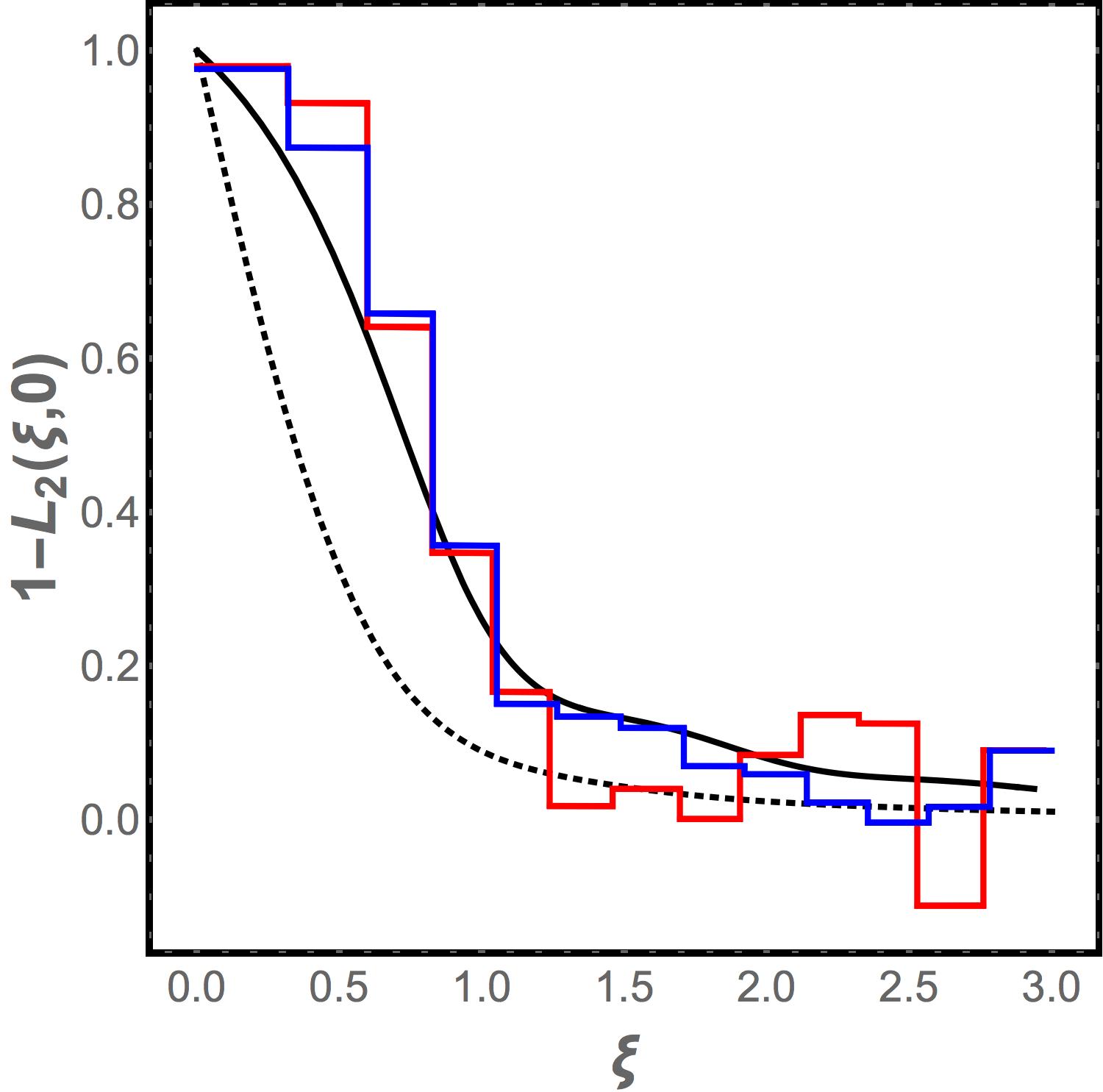}
\end{center}
\caption{Plot of the correlation function for the Laguerre ensemble (solid curve) and the GOE (dotted curve). The blue histogram
is the correlation function for an ensemble of 10,000 spectra drawn from the Laguerre ensemble generated as described in
Appendix \ref{sec:lsd}. The red histogram is the same but for a smaller ensemble of 1000 spectra. Comparison of the two
reveals that the departure of the numerical correlation from the exact analytic result is due in part to the finite
bind width (which is the same in both case) but in part also due to the finite size of the ensemble used to estimate the
correlation.}
\label{fig:lagcorrappendix}
\end{figure}

With these definitions established we can now write down the principal result of 
Nagao and Slevin~\cite{nagao} namely
\begin{eqnarray}
    1 - L_2 (\xi_1, \xi_2 ) & = & S (\xi_1, \xi_2) S ( \xi_2, \xi_1 )
        + \frac{1}{2} I (\xi_1, \xi_2 ) D ( \xi_2, \xi_1 ) 
        \nonumber \\
        & + & 
    \frac{1}{2} I (\xi_2, \xi_1 ) D (\xi_1,\xi_2 ).
    \label{eq:nsprincipal}
\end{eqnarray}
We are interested in $1 - L_2 (\xi, 0)$. For this special case we obtain the 
simplifications
\begin{eqnarray}
    S(\xi, 0 ) & = & \frac{\sqrt{2}}{r(x)}\left[\frac{1}{2} J_0 (x) + J_2 (x) \right], 
    \nonumber \\
    S( 0, \xi) & = & \frac{\sqrt{2}}{r(x)} \left[\frac{1}{2} J_0 (x) + \frac{1}{2} J_x (x) \right],
    \nonumber \\
    I (\xi, 0) & = & \frac{\sqrt{2}}{r(x)} \left[\int_0^x dt\; \frac{t}{2} {\cal C} (x, t) - \frac{1}{2}
    \right], \nonumber \\
    I (0, \xi) & = & \frac{\sqrt{2}}{r(x)} \left[\frac{1}{2} J_0 (x)\right],\nonumber \\
    D (\xi, 0) & = & - \frac{\sqrt{2}}{r(x)} \left[\frac{1}{6} J_2 (x) + \frac{1}{6} J_4 (x) \right], 
    \nonumber \\
    D( 0, \xi ) & = & - D(\xi, 0 ).
    \label{eq:simplifications}
\end{eqnarray}
Eqs. (\ref{eq:nsprincipal}) and (\ref{eq:simplifications}) are the principal results needed 
for our test of the random matrix model. 

Fig.\ref{fig:lagcorrappendix} shows a plot of $1-L_2(\xi,0)$ (solid black curve) as well as the
corresponding correlation for the GOE (dotted black curve) which is given by the expression in Eq. (\ref{eq:goetwopoint}). 
We have also computed the correlation function for the ensemble of ten thousand spectra drawn from the 
Laguerre ensemble as described in 
in Appendix \ref{sec:lsd}. This is shown as the blue histogram.
Since these spectra correspond to the very ensemble for which the solid curve represents
the exact solution the reason for any departure from the analytic result must be attributed to the 
finite width of the bins used in analyzing the numerical data and in the finite size of the
sample of spectra. This is confirmed by the red histogram which is based on a smaller ensemble
of one thousand spectra and shows markedly bigger departures from the analytic result especially for
larger $\xi$. Comparing the red histogram in Fig. \ref{fig:lagcorrappendix} to the histogram in 
Fig. \ref{fig:corrln} we see that the two agree about equally well with the analytic curve showing 
that the Laguerre ensemble prediction is indeed in excellent agreement with the numerical
data for the vibrational spectrum of jammed granular packs. 
\
In more detail the histograms are calculated as follows. We generate an ensemble of $M \times N$ 
random matrices $A_\alpha$ where $\alpha = 1, \ldots, \mu$ and $\mu$ is the number of realizations
in the ensemble. We then compute the eigenvalues of $A_\alpha^T A_\alpha$ and denote them
$\omega_{i\alpha}^2$ where $i = 1 \ldots N$. The range over which the eigenvalues lie is divided
into $\nu$ bins and the calculated spectra are binned. Denoting by $n_{i\alpha}$ the number of levels
in bin $i$ in realization $\alpha$ we then compute
\begin{equation}
    \langle n_i \rangle = \frac{1}{\mu} \sum_{\alpha = 1}^{\mu} n_{i \alpha}.
    \label{eq:meanoccupation}
\end{equation}
$\langle n_i \rangle$ is the unfolded width of the bin. The density of states correlator
between different bins is then estimated by 
\begin{equation}
    C_{ij} = 1 - \frac{1}{\mu} \sum_{\alpha=1}^{\mu} \frac{ n_{i \alpha} n_{j \alpha} }{ \langle n_i \rangle \langle n_j \rangle}.
    \label{eq:correlator}
\end{equation}
$C_{1j}$ then corresponds to $1 - L_2 (\xi, 0)$ over the interval
\begin{equation}
\sum_{i=1}^{j-1} \langle n_i \rangle < \xi < \sum_{i=1}^{j} \langle n_i \rangle. 
\label{eq:binwidth}
\end{equation}

\bibliography{rmt_spectrum}% Produces the bibliography via BibTeX.\

\begin{thebibliography}{999}

\bibitem{Coppersmith} C.-H. Liu et al, Science {\bf 269}, 513 (1995); S.N. Coppersmith et al, Phys. Rev. E {\bf 53}, 4673 (1996).

\bibitem{Cates} M.E. Cates, J.P. Wittmer, J-P. Bouchaud and P. Claudin, Phys. Rev. Lett. {\bf 81}, 1841 (1998).

\bibitem{MvHreview} M. van Hecke, J. Phys: Cond. Mat. {\bf 22}, 033101 (2009).

\bibitem{Bi} D. Bi, J. Zhang, B. Chakraborty and R.P. Behringer, Nature {\bf 480}, 355 (2011). 

\bibitem{Goldhirsch} I. Goldhirsch and G. Zanetti, Phys. Rev. Lett. {\bf 70}, 1619 (1993). 

\bibitem{vanNoije} T.P.C. vanNoije and M.H. Ernst, Granular Matter {\bf 1}, 57 (1998).

\bibitem{Reis} P.M. Reis, R.A. Ingale and M.D. Shattuck, Phys.Rev. E {\bf 75}, 051311 (2007).

\bibitem{Rouyer} F. Rouyer and N. Menon, Phys. Rev. Lett. {\bf 85}. 3676 (2000).

\bibitem{Hopkins} M.A. Hopkins and M.Y. Louge, Phys. Fluids A {\bf 3}, 47 (1991).

\bibitem{McNamara} S. McNamara and W.R. Young, Phys. Rev. E {\bf 50}, R28 (1994).

\bibitem{Brown} E. Brown and H.M. Jaeger, Rep. Prog. Phys. {\bf 77}, 046602 (2014).

\bibitem{LiuNagelCRCBook} A.J. Liu and S.R. Nagel eds., {\em Jamming and rheology: constrained dynamics on 
	microscopic and macroscopic scales.} (CRC Press, 2001.)

\bibitem{LiuNagel1998} A.J. Liu and S.R. Nagel, Nature {\bf 396}, 21 (1998).

\bibitem{OHern} C.S. O'Hern, L.E. Silbert, A.J. Liu and S.R. Nagel, Phys. Rev. E {\bf 68}, 011306 (2003).

\bibitem{Ellenbroek} W.G. Ellenbroek, E. Somfai, M. vanHecke, W.I.M. van Saarloos, Phys. Rev. Lett. {\bf 97}, 258001 (2006).

\bibitem{Wyart} M. Wyart, S.R. Nagel and T.A. Witten, Europhys. Lett. {\bf 72}, 486 (2005).

\bibitem{beltukov} Y.M. Beltukov, JETP Lett. {\bf 101}, 345 (2015).

\bibitem{mehta} M.L. Mehta, {\em Random Matrices.} (Academic Press, San Diego, 1991).

\bibitem{nagao} T. Nagao and K. Slevin, ``Laguerre ensembles of random matrices: Nonuniversal correlation
functions'', J. Math. Phys. {\bf 34}, 2317 (1993).  

\bibitem{Silbert} L.E. Silbert, A.J. Liu and S.R. Nagel, Phys. Rev. E {\bf 79}, 021308 (2009).

\bibitem{Nelson} Z. Zeravcic, W. van Saarloos and D.R. Nelson, Europhys. Lett. {\bf 83}, 44001 (2008). 

\bibitem{lubensky} C.L. Kane and T.C. Lubensky, Nat. Phys. {\bf 10}, 39 (2014).

\bibitem{calladine} C.R. Calladine, Int. J. Solids and Struct. {\bf 14}, 161 (1978) 

\bibitem{Forrester} P.J Forrester, {\em Log-Gases and Random Matrices} (Princeton University Press, Princeton, 2010).

\bibitem{Corey} K. VanderWerf, A. Boromand, M.D. Shattuck and C.S. O'Hern, Phys. Rev. Lett. {\bf 124}, 038004 (2020).

\bibitem{brezin} See, e.g., E. Br\'ezin and A. Zee, 
``Universality of the correlations between eigenvalues 
of large random matrices'', Nuclear Physics {\bf B402}, 613-627 (1993)

\bibitem{dyson} F.J. Dyson, Comm. Math. Phys. {\bf 19}, 245 (1970). 

\bibitem{Manning} M.L. Manning and A.J. Liu, Europhys. Lett. {\bf 109}, 36002 (2015). 

\bibitem{Middleton} E. Stanifer, P.K. Morse, A.A. Middleton and M.L. Manning, Phys. Rev. E {\bf 98}, 042908 (2018). 

\bibitem{Beltukov1} Y.M. Beltukov, V.I. Kozub and D.A. Parshin, Phys. Rev. B {\bf 87}, 134203 (2013).

\bibitem{Parisi} S. Franz, G. Parisi, P. Urbani and F. Zamponi, Proc. Nat. Acad. Sci. {\bf 112}, 14539 (2015).

\bibitem{Narayan} O. Narayan, Phys. Rev. E {\bf 63}, 010301 (2000).

\end{thebibliography}

\end{document}